\documentclass[twocolumn,showpacs,showkeys,preprintnumbers,amsmath,amssymb,floatfix]{revtex4}

\usepackage{graphicx}
\usepackage{dcolumn}
\usepackage{bm}

\newcommand{\beq}{\begin{equation}}
\newcommand{\eeq}{\end{equation}}
\newcommand{\bey}{\begin{eqnarray}}
\newcommand{\eey}{\end{eqnarray}}

\begin{document}

\title{ Finch-Skea star in (2+1) dimensions }

\author{Ayan Banerjee}
\email{ayan_7575@yahoo.co.in}
\affiliation {Department of Mathematics, Jadavpur University, Kolkata 700032, India.}
\author{Farook Rahaman}
\email{rahaman@iucaa.ernet.in}
\affiliation {Department of Mathematics, Jadavpur University, Kolkata 700032, India.}
\author{Kanti Jotania}
\email{kanti@iucaa.ernet.in}
\affiliation {Department of Physics, Faculty of Science, The M. S. University of Baroda, Vadodara 390 002, India.}
\author{Ranjan Sharma}
\email{rsharma@iucaa.ernet.in}
\affiliation{Department of Physics, P. D. Women's College, Jalpaiguri 735101, India.}
\author{Indrani Karar}
\email{indrani.karar08@gmail.com}
\affiliation{Department of Mathematics, Saroj Mohan Institute of Technology, Guptipara, West
Bengal, India.}

\date{\today}

\begin{abstract}
The Ba\~{n}ados, Teitelboim and Zanelli\cite{BTZ} solution corresponding to the exterior space-time of a black hole in $(2+1)$
 dimensions has been found to be very useful to understand various aspects relating to the gravitational field of
 a black hole. We present here a class of interior solutions corresponding to the BTZ exterior by making use of a model
 presented by Finch and Skea\cite{Finch} which was earlier found to be relevant for the description realistic stars in
  $(3+1)$ dimensions. We show physical viability of the model in lower dimensions as well.
\end{abstract}
\keywords{Einstein's field equations; Stellar equilibrium; $(2+1)$
dimensional gravity, BTZ solution.}
\pacs{04.40.Nr, 04.20.Jb,
04.20.Dw}
\maketitle

\section{Introduction}
Investigations in lower dimensional gravity play a crucial role towards our understanding of various aspects of gravity. Some of the fundamental problems in $4$ dimensional gravitational analyses become much simpler in $(2+1)$ dimensions for obvious reasons. In the studies of black holes, a lower dimensional analysis has often been preferred to understand various issues which are otherwise difficult to resolve in conventional dimensions. For example, Ba\~{n}ados, Teitelboim and Zanelli\cite{BTZ}    (henceforth BTZ), in the presence of a negative cosmological constant, have obtained an analytic solution representing the exterior gravitational field of a black hole in $(2+1)$ dimensions which has opened up the possibility of investigating many interesting features of black holes.

The objective of the present work is to obtain a class of interior solutions corresponding to the BTZ exterior
 metric describing a static circularly symmetric star in equilibrium. Collapse of a dust cloud in $(2+1)$
 dimensions leading to a black hole was analyzed by Mann and Ross\cite{Mann}. Considering the collapse of a circularly symmetric anisotropic fluid, a self-similar solution in $(2+1)$ dimensions was obtained by Martins {\em et al}\cite{Martins}. Cruz and Zanelli\cite{Cruz} obtained an interior solution for an incompressible fluid in $(2+1)$ dimensions and investigated the bound on the maximum allowed mass of the resultant configuration. By assuming a particular density profile, a class of interior solutions corresponding to BTZ exterior was provided by Cruz {\em et al}\cite{Cruz2}. Paulo M. S$\acute{a}$\cite{Paulo} proposed an interior solution corresponding to the BTZ exterior where a polytropic type equation of state (EOS) was assumed. Sharma {\em et al}\cite{Sharma} assumed a particular form of the mass function to obtain a new class of interior solutions corresponding to the BTZ exterior. At the back drop of such varied developments, Garc\'{i}ýa and Campuzano\cite{Garc} provided a general methodology to construct all possible types of solutions in $(2+1)$ dimensions for a static circularly symmetric perfect fluid source. The exact analytic form of the metric potentials, in this formulation, can be obtained for any arbitrary choice of the density profile or EOS of the matter content of the fluid source. In particular, they presented a solution corresponding to a static circularly symmetric perfect fluid source having constant energy density, which (in the presence of a cosmological constant) might be considered as analogous to the incompressible Schwarzschild interior solution in $(3+1)$ dimensions. Since determination of the exact analytic form of the solution in this formalism requires knowledge about the EOS of the composition or the radial dependence of energy-density, we find it worthwhile to adopt an alternative method where the right hand side of the Einstein's field equations ($T_{ij}$) will be governed by the geometry of the associated space-time ($G_{ij}$). In this vein, following Finch and Skea\cite{Finch}, we use the ansatz for one of the metric functions ($g_{rr}$) describing the background space-time and determine the remaining metric function $g_{tt}$ by solving the relevant Einstein's field equations. In $(3+1)$ dimensions, the Finch and Skea\cite{Finch} ansatz has been found to be useful to develop physically acceptable models capable of describing realistic stars. The physical $3$-space of the back ground space-time, when embedded in a $4$-dimensional Euclidean space, represents a $3$-paraboloid which is a departure from the $3$-spherical geometry \cite{Tik1},\cite{Tik2}. Similarly, in $(2+1)$ dimensions, though the space-time remains circularly symmetric, the $t=constant$ hyper-surface of the associated background space-time becomes a parabola rather than a circle. In the present work, we show that by making use of the Finch and Skea\cite{Finch} ansatz in $(2+1)$ dimensions, it is possible to generate physically acceptable interior solutions corresponding to the BTZ exterior metric. Though the class of solutions obtained here may be regarded as a particular case of the general formalism mentioned in Ref.~\cite{Garc}, the two approaches are complementary in nature. Our geometric approach may open up the possibility of analyzing the role of space-time geometry on the dynamics of a collapsing system which, however, is a matter of further investigation. 

\section{Interior space-time}
We write the line element for the interior space-time of a static circularly symmetric Finch and Skea\cite{Finch} type star in
 $(2+1)$ dimensions as
\begin{equation}
ds_{-}^2 = -e^{2\nu(r)}dt^2 + \left(1+\frac{r^2}{R^2}\right)dr^2 +r^2 d\theta^2,\label{eq1}
\end{equation}
where $R$ is a curvature parameter governing the geometry of the background space-time. Note that the $t=constant$ hyper-surface
 of the metric (\ref{eq1}) is parabolic in nature. We assume a perfect fluid type matter distribution for the
  interior of the star and accordingly write the energy-momentum tensor in the form
\begin{equation}
T_{ij} = (\rho + p)u_{i}u_{j} + p g_{ij},\label{eq2}
\end{equation}
where $u^i = e^{-\nu}\delta_t^i$ is 3-velocity of the fluid. In (\ref{eq2}), $\rho$ and $p$ represent the energy density and isotropic pressure of the matter distribution of the star, respectively.

The Einstein's field equations with a negative cosmological constant ($\Lambda < 0$) are then obtained as (we assume $G = c = 1$)
\begin{eqnarray}
2\pi\rho+\Lambda &=& \frac{1}{R^2}\left(1+\frac{r^2}{R^2}\right)^{-2},\label{eq3}\\
2\pi p - \Lambda &=& \frac{\nu^{\prime}}{r}\left(1+\frac{r^2}{R^2}\right)^{-1},\label{eq4}\\
2\pi p - \Lambda &=& \left(1+\frac{r^2}{R^2}\right)^{-1}\left({\nu^{\prime}}^2+\nu^{\prime\prime}-\frac{\nu^{\prime}}{r^2+R^2}r\right),\label{eq5}
\end{eqnarray}
where a '$\prime$' denotes the differentiation with respect to $r$.

The mass function $m(r)$ of the star may be obtained by integrating Eq.~(\ref{eq3}) which yields
\begin{equation}
m(r) = \int_{0}^{r}2\pi\rho rdr = \frac{r^2}{2\left(r^2+R^2\right)}-\frac{\Lambda r^2}{2}.\label{eq6}
\end{equation}
The above equation shows that $m(r) =0$ at $r=0$ and is positive for any finite radius since $\Lambda$ has been assumed to be negative.

To determine the metric potential $\nu(r)$, we combine Eqs.~(\ref{eq4})-(\ref{eq5}) which yields
\begin{equation}
\frac{\nu^{\prime}}{r} = {\nu^{\prime}}^2+{\nu^{\prime\prime}}-\frac{\nu^{\prime}r}{r^2+R^2}.\label{eq7}
\end{equation}
Integrating Eq.~(\ref{eq7}), we get
\begin{equation}
\frac{\nu^{\prime}}{r} = \frac{\sqrt{r^2+R^2}}{\frac{1}{3}(r^2+R^2)^{3/2}+B},\label{eq8}
\end {equation}
where $B$ is an integration constant. Integration of Eq.~({\ref{eq8}) helps us to determine the unknown metric function $\nu(r)$ in the form
\begin{equation}
\nu(r)= \ln\left[C\left\{\frac{1}{3}(r^2+R^2)^{3/2} + B\right\}\right],\label{eq9}
\end{equation}
where $C$ is an integration constant.

From Eq.~(\ref{eq4}), the isotropic pressure is then obtained as
\begin{equation}
p = \frac{1}{2\pi}\left(1+\frac{r^2}{R^2}\right)^{-1}\left[\frac{\sqrt{r^2+R^2}}{\frac{1}{3}(r^2+R^2)^{3/2}+B}\right]+\frac{\Lambda}{2\pi}.\label{eq10}
\end {equation}
It is to be noted here that for obtaining an analytic solution, the technique adopted here is equivalent to prescribing a density profile of the form (\ref{eq3}) in the formulation discussed in Ref.~\cite{Garc}. 

Now, in $(3+1)$ dimensions the metric (\ref{eq1}) gets the form
\begin{equation}
ds_{-}^2 = -e^{2\nu(r)}dt^2 + \left(1+\frac{r^2}{R^2}\right)dr^2 +r^2 (d\theta^2 + \sin^2\theta d\phi^2).\label{eq11}
\end{equation}
Assuming that the metric (\ref{eq11}) describes the space-time of a perfect fluid with energy-momentum tensor of the form (\ref{eq2}), one can utilize the Einstein's field equations to write the pressure isotropy condition in the form
\begin{equation}
\frac{\nu^{\prime}}{r} = {\nu^{\prime}}^2+{\nu^{\prime\prime}}-\frac{\nu^{\prime}r}{r^2+R^2} +\frac{r^2}{R^2(r^2+R^2)}.\label{eq12}
\end{equation}
It is interesting to note that Eqs.~(\ref{eq7}) and (\ref{eq12}) are almost identical except for the last term appearing on the right hand side of Eq.~(\ref{eq12}). The solution to Eq.~(\ref{eq12}), as provided by Finch and Skea\cite{Finch}, is given by
\begin{eqnarray}
e^{\nu} &=& \left(D - A\sqrt{1+\frac{r^2}{R^2}}\right)\cos\sqrt{1+\frac{r^2}{R^2}} + \nonumber \\
&& \left(D\sqrt{1+\frac{r^2}{R^2}} + A\right)\sin\sqrt{1+\frac{r^2}{R^2}},\label{eq13}
\end{eqnarray}
where, $A$ and $D$ are constants whose values are determined by using the boundary conditions across the boundary surface joining the interior and the exterior region.  The BTZ metric is replaced by the Schwarzschild metric in $(3+1)$ dimensional model developed by Finch and Skea\cite{Finch}. In fact, any physically motivated appropriate exterior space-time region may be taken up in this extrapolation procedure to understand the nature of the physical parameters of the configuration. However, an attempt to obtain the corresponding solution in $(2+1)$ dimensions by a dimensional reduction method will be of little significance as it is well known that vacuum solutions corresponding to the Einstein's field equations in $(2+1)$ dimensions are necessarily flat. A meaningful approach in such a situation would be to consider an anti-de Sitter type exterior region. However, since we are interested in obtaining an interior solution corresponding to the BTZ exterior metric, in the following section we write down the appropriate junction conditions and analyze the physical behaviour of the model accordingly.

\section{Exterior space-time and matching conditions}
We assume that the exterior space-time of the configuration is described by the BTZ  metric given by
\begin{equation}
ds{+}^{2} = -\left(-M_0 - \Lambda{r^2} \right)dt^2 + \left(-M_0 -\Lambda{r^2} \right)^{-1}dr^2 + r^2d\theta^2,\label{eq14}
\end{equation}
where $M_0$ corresponds to the conserved charge associated with the asymptotic invariance under time displacements.
If $a$ is assumed to be the radius of the star, continuity
of the metric functions $g_{tt}$ and $g_{rr}$  across the boundary yields the following results:
\begin{eqnarray}
\left[C\left\{\frac{1}{3}(a^2+R^2)^{3/2}+B\right\}\right]^{2} = -M_0-\Lambda a^2,\label{eq15}\\
\left(1+\frac{a^2}{R^2}\right)^{-1} = -M_0-\Lambda
a^2,\label{eq16}
\end{eqnarray}
Since pressure must be zero at the boundary of the star $r=a$, from Eq.~(\ref{eq10}), we get
\begin{equation}
\left[\frac{\sqrt{a^2+R^2}}{\frac{1}{3}(a^2+R^2)^{3/2}+B}\right] = -\Lambda
\left(1+\frac{a^2}{R^2}\right).\label{eq17}
\end{equation}
We solve Eqs.~(\ref{eq15})-(\ref{eq17}) simultaneously and determine the constants as
\begin{eqnarray}
B &=& \left(\frac{a}{-\Lambda}\right) \left[\frac{(-\Lambda a^2-M_0)}{\sqrt{(1+M_0+ \Lambda a^2)}}\right]
\nonumber\\
&&~~~~~~ - \frac{a^3}{3(1+M_0+\Lambda a^2)^{\frac{3}{2}}} ,\label{eq18}\\
C &=& \left(\frac{-\Lambda}{a}\right)\frac{1}{\sqrt{(-\Lambda a^2-M_0)(1+M_0+\Lambda a^2)}},\label{eq19}\\
R & =&\sqrt{\frac{a^2(-\Lambda a^2-M_0)}{1+M_0+\Lambda a^2}}.\label{eq20}
\end{eqnarray}
Using Eqs.~(\ref{eq6}) and (\ref{eq16}), we express the total mass $M(a)$ bounded within the boundary surface $a$ as
\begin{equation}
M(a) = \frac{a^2}{2\left(a^2+R^2\right)}-\frac{\Lambda a^2}{2} = \frac{1}{2}(1+M_0).\label{eq21}
\end{equation}
Now, for appropriate signature of the metric functions we must have $ - M_0 -\Lambda a^2 > 0$, i.e., $ a > \sqrt{-M_0/\Lambda}$ which puts a lower bound on $M_0$ such that $M_0 = 0$. The values of $\Lambda$, $M_0$ and $a$ should be so chosen that the above condition is satisfied. Noting that $M_0 < -\Lambda a^2$, for a given set of values of $\Lambda$ and $a$, the upper limit on the mass may be written as
\begin{equation}
M(a)_{max} < \frac{1}{2}(1 -\Lambda a^2).\label{eq22}
\end{equation}
From Eq.~(\ref{eq21}), the compactness of the stellar configuration is obtained as
\begin{equation}
\mathfrak{u}=\frac{m(r)}{r}|_{r=a}  =
\frac{a}{2}\left[\frac{1}{a^2+R^2} -\Lambda\right].\label{eq23}
\end{equation}
Accordingly, the surface redshift parameter $z_{s}$ gets the form
\begin{equation}
z_{s} = \left[1-a\left(\frac{1}{a^2 + R^2} -\Lambda\right)\right]^{-\frac{1}{2}}-1.\label{eq24}
\end{equation}

\subsection{Physical acceptability of the model}
A physically acceptable model must comply with certain regularity conditions. The density and pressure should be
 positive at the interior of the star and both should decrease monotonically from the centre towards the boundary, i.e.,
  the following conditions should be satisfied: (i) $\rho \geq 0$, $p \geq 0$, (ii) $\frac{d\rho}{dr} < 0$ and $\frac{dp}{dr} < 0$.
  Moreover, for microscopic stability we must have $ \frac{dp}{d\rho} > 0$. In our model, from Eqs.~(\ref{eq3}) and (\ref{eq10}),
  the central density and central pressure are obtained as
\begin{eqnarray}
\rho_{0} &=& \frac{1}{2\pi}\left(\frac{1}{R^2} -\Lambda\right),\label{eq25}\\
p_{0} &=& \frac{1}{2\pi}\left(\frac{R}{\frac{1}{3}R^3+B}+ \Lambda\right).\label{eq26}
\end{eqnarray}
Therefore, both $\rho_{0}$ and $p_{0}$ will be positive if the condition
\begin{equation}
\frac{R}{R^3/3 + B} > -\Lambda,\label{eq27}
\end{equation}
is satisfied. From Eqs.~(\ref{eq3}) and (\ref{eq10}), we also obtain
\begin{eqnarray}
\frac{d\rho}{dr} &=& -\frac{2rR^2}{\pi {\left(r^2+R^2\right)^3}},\label{eq28}\\
\frac{dp}{dr} &=& -\frac{r}{2\pi}\left(1+\frac{r^2}{R^2}\right)^{-1}\left[
\frac{\frac{2}{3}(r^2+R^2)-\frac{B}{\sqrt{r^2+R^2}}}
{[{\frac{1}{3}(r^2+R^2)^{3/2}+B}]^2}\right. \nonumber\\
&& \left. + \frac{2}{\sqrt{r^2+R^2}{[{\frac{1}{3}(r^2+R^2)^{3/2}+B}]}}\right]. \label{eq29}
\end{eqnarray}
Obviously, $\frac{d\rho}{dr} = 0 = \frac{dp}{dr}$ at $r = 0$. It can be shown that, for appropriate choices of the model parameters, the density and
pressure are decreasing functions of the radial parameters $r$.

From Eqs.~(\ref{eq28})-(\ref{eq29}), we have
\begin{eqnarray}
\frac{dp}{d\rho} = \frac{(r^2+R^2)^2}{4}\left[\frac{\frac{2}{3}(r^2+R^2)-\frac{B}{\sqrt{r^2+R^2}}}
{[{\frac{1}{3}(r^2+R^2)^{3/2}+B}]^2}\right. \nonumber\\
\left. + \frac{2}{\sqrt{r^2+R^2}{[{\frac{1}{3}(r^2+R^2)^{3/2}+B}]}}\right], \label{eq30}
\end{eqnarray}
which should be positive inside the star.

To check that the model complies with the above mentioned requirements, we assume $a =3$ and $M_0 = 0.2$. In the anti de-Sitter space-time the cosmological constant is assumed to be negative and, accordingly, we choose $\Lambda = -0.04$ (see Ref.~\cite{Fernando,Astefanesei}). Eqs.~(\ref{eq18})-(\ref{eq20}), then determine the constants as $R = 1.3093$, $B =
1.4028$ and $C = 0.03637$. From Eq.~(\ref{eq21}), the total mass is obtained as $M(a) = 0.6$. The compactness is found to be $M(a)/a = 0.2$ and the corresponding surface red-shift is obtained as $z_s = 0.291$. In Fig.~(\ref{fig:1}) and (\ref{fig:2}), we have plotted variations of the energy density and pressure which clearly indicate their regular behaviour at the stellar interior. In Fig.~(\ref{fig:3}), the maximum permissible masses for given radii have been shown for different choices of the constant $\Lambda$. The plot indicates that, for a given radius, compactness can be increased by lowering the value of the constant $\Lambda$. This has been shown in a tabular form in Table~\ref{tab:table1}. In Fig.~(\ref{fig:4}), $dp/d\rho$ has been plotted against $r$ which shows that $dp/d\rho > 0$ at all interior points of the star.

\begin{table}
\caption{\label{tab:table1} Maximum permissible mass for a given radius at different $\Lambda$.}
\begin{ruledtabular}
\begin{tabular}{lcr}
$a$ (km) & $\Lambda$ (km$^{-2}$) & $M_{max}$ ($M_{\odot}$)\\
\hline
10 & -0.04 & 1.69\\
10 & -0.1 & 3.73\\
10 & -0.2 & 7.12\\
\end{tabular}
\end{ruledtabular}
\end{table}

\begin{figure}
\centering
\includegraphics[scale=.6]{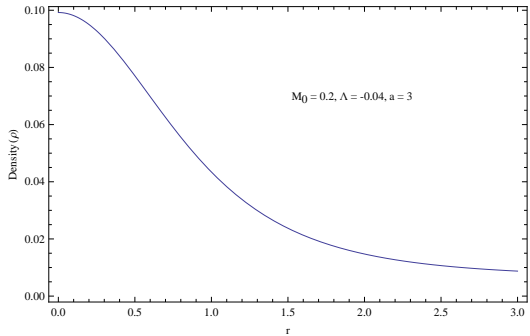}
\caption{Radial variation of energy density ($\rho$). }
    \label{fig:1}
\end{figure}

\begin{figure}
\centering
\includegraphics[scale=.6]{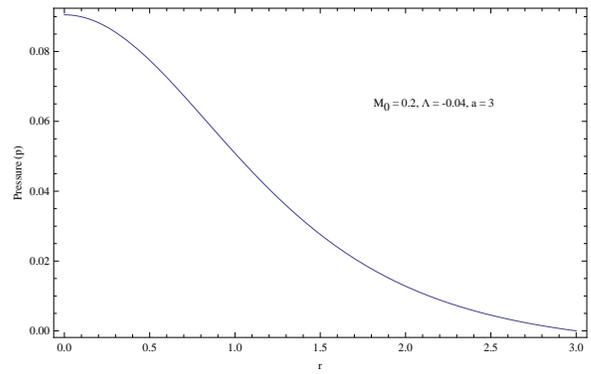}
\caption{Radial variation of isotropic pressure ($p$).}
    \label{fig:2}
\end{figure}

\begin{figure}
\centering
\includegraphics[scale=.6]{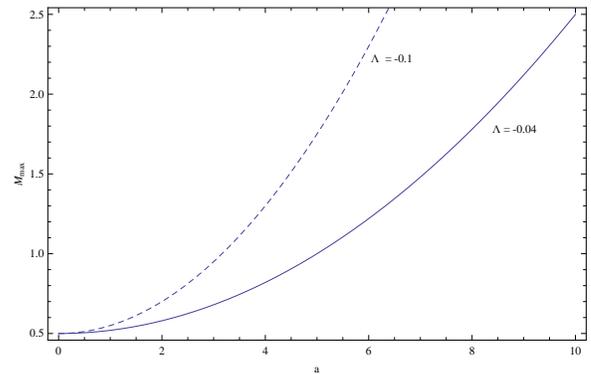}
\caption{$M_{max}$ plotted against $a$ for different $\Lambda$.}
    \label{fig:3}
\end{figure}

\begin{figure}
\centering
\includegraphics[scale=.6]{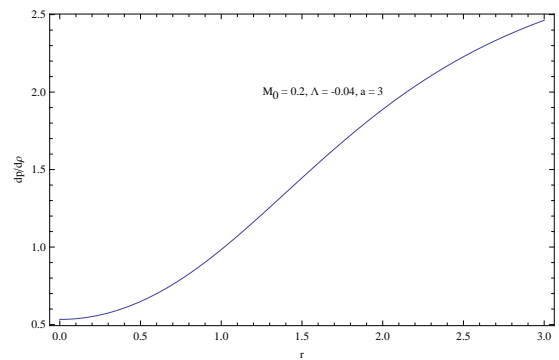}
\caption{$\frac{dp}{d\rho}$ plotted against $r$.}
    \label{fig:4}
\end{figure}

\section{Discussions}
By applying the Finch and Skea\cite{Finch} ansatz in $(2+1)$ dimensions, we have obtained a class of interior solutions corresponding to the exterior space-time described by the Ba\~{n}ados, Teitelboim and Zanelli\cite{BTZ} solution. Based on physical grounds, we have obtained bounds on the model parameters and demonstrated that by suitably fixing the model parameters, it is possible to develop viable models for the interior space-time corresponding to the BTZ exterior. We have obtained an upper bound on the mass content and also analyzed the impact of cosmological constant on the compactness of the system. The geometric approach adopted here may be considered as a complementary to the method prescribed in Ref.~\cite{Garc} and provides an opportunity to analyze the impact of space-time geometry on the dynamics of a collapsing system. To achieve this goal, we intend to utilize the static solution to generate models of evolving systems in $(2+1)$ dimensions. Work is in progress in this direction and will be reported elsewhere.  

\subsection*{Acknowledgments}
We would like to thank the anonymous referee for his useful suggestions. FR, KJ and RS are thankful to the Inter University Centre for Astronomy and Astrophysics (IUCAA), India, where a part of this work was carried out under its visiting research associateship programme. FR is also grateful to UGC, India, for financial support under its Research Award Scheme.

\end{document}